\documentstyle[12pt]{article}
\newcommand{\bea}{\begin{eqnarray}}
\newcommand{\be}{\begin{equation}}
\newcommand{\eea}{\end{eqnarray}}
\newcommand{\ee}{\end{equation}}
\def\nn{\nonumber}

\def\a{\alpha}
\def\b{\beta}

\def\d{\delta}
\def\e{\epsilon}

\def\l{\lambda}
\def\m{\mu}

\def\o{\omega}
\def\p{\pi}

\def\th{\theta}

\def\t{\tau}

\def\x{\xi}

\def\J{\Psi}
\def\L{\Lambda}

\def\cc{{\cal C}}

\def\ch{{\cal H}}

\def\cm{{\cal M}}

\def\co{{\cal O}}

\def\car{{\cal R}}

\def\cz{{\cal Z}}

\begin{document}
\pagestyle{plain} \large \noindent . \vspace*{5mm} \\
CGPG-95/1-1\\January 1995\\revised March 1995
\begin{center} \LARGE \bf \vspace*{5mm}
Large diffeomorphisms in (2+1)-quantum gravity on the torus\\
\vspace*{10mm} \large \bf
 Peter Peld\'{a}n
\footnote{Email address: peldan@phys.psu.edu}\\
\vspace*{5mm} \large
 Center for Gravitational Physics and Geometry\\
Physics Department\\
The Pennsylvania State University, University Park, PA 16802, USA\\

\vspace*{10mm} \Large \bf
Abstract\\
\end{center} \normalsize
The issue of how to deal with the modular transformations -- large
diffeomorphisms --
in (2+1)-quantum gravity on the torus is discussed.
I study the Chern-Simons/connection representation and show that the behavior
of the
modular transformations on the reduced configuration space is so bad that it
 is
possible to rule out all finite dimensional unitary representations of the
modular
group on the Hilbert space of $L^2$-functions on the reduced configuration
space. Furthermore, by assuming piecewise continuity for a dense subset of
the vectors in any Hilbert space based on the space of complex valued
functions on the reduced configuration space, it is shown that finite
dimensional representations are excluded no matter what inner-product we
define in this vector space.
A brief discussion of the loop- and
ADM-representations is also included.
\section{Introduction}
During the last six years there has been an increasing interest in the study of
(2+1)-dimensional quantum gravity, and specifically "(2+1)-gravity on the
torus".
However, although one now can say that gravity has been successfully quantized
in
(2+1)-dimensions with the use of several different formulations as well as
methods,
there are still not many physically interesting questions that have been
answered
definitely. (Maybe because there are no definite answers.) Examples of such
interesting questions are; 1) Is area or length quantized in (2+1)-quantum
gravity? 2)
What happens to the classical spacetime singularity upon quantization? 3) What
is a
complete set of observables in quantum gravity, and what is the Hilbert space
upon
which these observables act irreducibly?

Furthermore, although most workers in the field seem to agree on how to handle
the
local symmetries -- in principle -- there still seems to be quite a lot of
disagreement of how to handle the global symmetries of the theory. Some authors
claim
that the quantum theory must be invariant under those symmetries as well
\cite{carlip1}, \cite{carlip2}, \cite{menotti}, \cite{hosoya}
while others believe it is enough to have the
wavefunctions
transforming under a finite-dimensional unitary
representation of the symmetry group
\cite{arlen}, \cite{giulini}.
A third alternative is given in \cite{ashtekar1} \cite{carlip3} where one
basically
claims that the
global
symmetries should almost not restrict the quantum theory at all; it is enough
if the
wavefunctions transform under any unitary representation of the symmetry group.
A
fourth group of papers do not explicitly discuss how to deal with the global
symmetries \cite{ezawa}, \cite{louko}, \cite{witten}, \cite{unruh},
\cite{marolf}, \cite{regge}.\\ \\
Now, to better understand why different researchers in the field have such a
different opinion about the treatment of the global symmetries in quantum
gravity, we
may study the conventional treatment of global symmetries in standard quantum
mechanics. Then we need to distinguish between the two cases: 1) we have a
global
symmetry that commutes with all the observables in the theory, 2) or the global
symmetry does not commute with all observables.

In the first case, we get a so called
super selection rule. That is, if we assume that our Hilbert space may be
written as a
direct sum (integral) of subspaces left invariant by the symmetry, there is no
way we
can distinguish between states that only differ by the superposition
coefficients
between these subspaces. Thus, the conclusion must be that every subspace
defines a
separate quantum representation, and since they do not communicate there is no
reason
to treat them together; we might as well study each sector separately. Often,
these
subspaces are of the same dimension -- infinite dimensional --  and can be
considered
isomorphic.

One of the simplest example of such a situation is the parity symmetry
in the Hilbert space of $L^2(\car )$-functions. Suppose that all our
observables are
invariant under parity. Then there is no way to distinguish between the
wavefunctions $\J _1(x):=\J _s(x)+\J _a(x)$ and $\J _2(x):=\J _s(x)-\J _a(x)$.
Here
$\J _s(x)=\J _s(-x)$ and $\J _a(x)=-\J _a(-x)$. The wavefunctions $\J _s(x)$
and
$\J _a(x)$ transform under different unitary irreducible representations and
they do
not communicate. Thus, the even and odd sectors define two isomorphic quantum
theories.

A similar situation appears in $SU(N)$ Yang-Mills theory \cite{theta}.
Here the global symmetries are the large $SU(N)$ gauge transformations. Since
the
physical observables in this theory normally are defined in terms of gauge
invariant
functions of the electric and magnetic field, the observables are automatically
invariant under the large gauge transformations as well. We get super selection
rules. By assuming the gauge transformations to fall off to the
identity-transformation at spatial infinity, one may show that the group of
large
transformations modulo the small ones is isomorphic to the group of integers.
Thus,
the only irreducible representations are the one-dimensional ones and they are
given
by $D(g)=\exp (i \th )$ where $g$ is the generator of the group and $\th$ is a
parameter labeling the different representations. Hence, we see that we get an
infinite number of irreducible representations and since they do not
communicate,
there is no way to distinguish between states that only differ in the
superposition
between the representations; we could study one sector at the time.

Normally, in these cases,
one implicitly assumes that every unitary representation can be
written as a direct sum (integral) of unitary irreducible representations of
the
symmetry group. This is not always the case. There is a classification of group
representations
into type I, II and III \cite {mackey}, and it is only for type I
representations
that this
decomposition always can be performed. For type II and III it may happen that
it is
impossible to write a given representation as a direct sum (integral) of
irreducible
representations, or, if a decomposition exists, it may not be unique. The
relevant
group of large diffeomorphisms for quantum gravity on the torus, $PSL(2,\cz )$,
allows
representations of type II\footnote{Amongst the separable locally compact
groups
which are known to have only type I representations are the compact groups, the
commutative groups, the connected semi-simple Lie-groups and the connected real
algebraic groups. However there are many examples of connected solvable Lie
groups
which have non-type I representations. It is known that the regular
representation of
a discrete group is of type II whenever the subgroup generated by the finite
conjugate
classes has infinite index \cite{mackey}.}
, and there is therefore no reason to believe that the
Hilbert space should
be totally decomposable into irreducible subspaces.

In case 2 -- where the global symmetry does not commute with all observables --
we
get a similar situation as for case 1. Here, however, since the observables do
not
commute with the symmetry, we can distinguish between wavefunctions that only
differ
by a symmetry transformation. Thus, the situation will be that our Hilbert
space may
be decomposable into a direct sum (integral) of irreducible subspaces, but
since the
sectors now do communicate we cannot study each sector separately; we need all
related
sectors. However, note here as well that if our representation is not of type
I, there
is no guarantee that the Hilbert space is decomposable into irreducible
pieces.\\ \\
Thus, we may now understand one reason for disagreement about the treatment of
global
diffeomorphisms in quantum gravity; if one believes that all physical
observables
should be invariant under all diffeomorphisms, we get case 1, and if one only
requires
the physical observables to be invariant under the small diffeomorphisms, we
get case
2. Consequently, the situation seems rather clear in both cases whenever the
representation of the symmetry
group is of type I. In these cases one could just follow the standard text book
treatment.

Now, what are the arguments to support these two different
viewpoints? It seems that if we take seriously the fact that all
diffeomorphisms -- large or small -- are pure gauge and we can
therefore never physically distinguish between systems related by
diffeomorphisms,
we are inevitably led to
the conclusion that all our physical observables must be
invariant under all diffeomorphisms.

The other viewpoint -- that observables do not have to commute
with the large diffeomorphisms or gauge transformations -- follows directly
from the
conventional Dirac-definition of
observables in constrained systems; a physical observable is a
function on the phase space that commutes with all first class
constraints.

As I see it, there are huge problems with both these viewpoints
and we cannot categorically say that we always should treat all
diffeomorphisms in a unified way. For instance, by always
demanding invariance under all diffeomorphisms, we would have to
require wavefunctions for an asymptotically flat spacetime to
transform trivially under diffeomorphisms that tend to Poincare
transformations at infinity. Most physicists would probably agree
that this is a too restricted quantum theory. The problem with
the other viewpoint is that we now allow objects to be called
physical observables although we in principle never can set up a
measuring-apparatus to measure them. Normally, in standard
quantum mechanics, we say that not all self adjoint operators can
be considered as physical observables; a physical observable
needs to be in principle measurable. Take for instance the quantum
theory of identical particles. We have a global symmetry;
permutations of the particles. According to the Dirac-definition
of a physical observable, there is nothing wrong with observables
that do not commute with the permutation operator. However, we
know that if our particles are truly indistinguishable, we can
never set up a measuring device to measure any quantity that is
not invariant under permutations of the particles. We define a
physical observable to be a self adjoint operator that commutes
with the permutation operator \cite{qm}.

Is there any way out of this dilemma? Yes, I believe that if we
stop trying to quantize gravity by applying standard
quantization methods to the entire spacetime and instead start
studying quantum gravity for bounded regions of spacetime, this
problem will be naturally solved. That is, when we are trying to
use standard arguments from conventional quantum mechanics in a
closed quantum mechanical system; quantum gravity for the the
entire spacetime, we are really applying our quantization
procedures to a situation outside the domain where our quantization rules
applies.
Quantum mechanics -- in its current form -- is constructed to be
applied to physical systems consisting of a classical observer
together with the quantum system. If we try to quantize closed
systems without classical observers, we have actually no real
guidance from conventional quantum mechanics and we are
therefore free to choose a definition of what we mean by physical observables.
At the
end, experiments will tell.

However, if we instead study bounded regions in spacetime, and
quantize only the interior of the region while keeping the
outside classical, we could probably carry over most arguments
and procedures from standard quantum mechanics.

In short, I believe the situation would be the following: all
diffeomorphisms that leave the boundary invariant
(diffeomorphisms that goes to the identity transformation at the
boundary) should be
considered pure gauge and should therefore define superselection
rules for the wavefunctions. However, for diffeomorphisms that
do not leave the boundary invariant, we just get ordinary
symmetries and there is no {\it a priori} reason to restrict the
quantum theory to any specific representation. This other group
of diffeomorphisms -- diffreomorphisms that do not go to the
identity at the boundary -- will therefore be treated analogous to Galilean
transformations in classical Newtonian mechanics and Poincare
transformations in special relativity.\\ \\
Thus my opinion is that conventional quantum mechanics does not
really give us any instructions of how to handle quantum
mechanics for closed systems without classical observers. Or, it
may even be that it does not make sense to talk about closed
quantum systems in the absence of classical observers. And, if
we still want to study such quantum theories, we are
free to choose a definition of a physical observable ourselves, and therefore
it
makes sense to study both approaches: 1) observables must
commute with all unphysical transformations, 2) observables just
need to commute with small transformations.\\ \\
In this paper, I choose to study the first approach. I require
all observables to be invariant under the full group of
diffeomorphisms on the torus, and therefore -- according to
standard arguments for superselection rules in quantum mechanics
\cite{qm} -- I should only study the one-dimensional
representations. (Just to be safe I will keep the representation
slightly more general; finite dimensional, unitary, and
 irreducible.) I will show that in the connection representation
of quantum gravity on the torus, the quantum theory that follows
from this approach becomes completely trivial; the Hilbert space
is at most one-dimensional if we require the wavefunctions to both be
piecewise continuous somewhere on the reduced configuration
space and also to transform under a finite-dimensional representation of
$PSL(2,\cz )$.
I also give a brief discussion of
the case of the loop-representation and the reduced ADM-phase space
quantization. There, the
irreducible sectors seems to be non-trivial.

\section{The Chern-Simons/connection approach}
In the Chern-Simons/connection approach to Einstein gravity, the starting point
is the
Chern-Simons action
\be S_{CS}=\int d^3x\e ^{\a \b \d}Tr\left (A_\a\partial _\b A_\d +
\frac{1}{3}A_\a A_\b
A_\d \right ) \ee
or the Hilbert-Palatini action
\be S_{HP}=\int d^3x \e ^{\a \b \d}e_\a ^IR_{\b \d I}(\o) \ee
where $A_\a$ is an $ISO(1,2)$ valued connection, $Tr$ is an non-degenerate
invariant bilinear form on the Lie-algebra of
$ISO(1,2)$, $e_\a^I$ is the triad field, and $\o _\a ^I$ is an $SO(1,2)$
connection.
These two actions can be shown to be equivalent provided $A_\a =e_\a ^IP_I + \o
_\a
^IJ_I$, where $P_I$ are the generators of three-dimensional translations and
$J_I$ are
the generators of $so(1,2)$. See \cite{witten} for details. To quantize this
formulation, one may proceed as follows: 1) Do a (2+1)-decomposition to
the Hamiltonian formulation. 2) Identify the phase space variables, and the
first class
constraints. 3) Solve the constraints and gauge fix the local symmetries. 4)
Identify the remaining physical degrees of freedom and the Hamiltonian. 5)
Quantize.\\
I will not discuss these steps in detail here. See {\it e.g} \cite{carlip1},
\cite{carlip2}, \cite{marolf}, \cite{ashtekar1}
for details. In short, one notices that the constraints tell us that the
reduced
configuration space consists of gauge equivalent classes of flat $so(1,2)$
connections, and since a flat connection is completely determined -- up to
gauge
transformations -- by its holonomies around the non-contractible loops on the
hypersurface, one may parameterize the reduced configuration space by two
commuting
$SO(1,2)$ element. They have to commute since they are suppose to give a
representation of the fundamental group of the torus, which is $\p _1(T^2)\sim
\cz
\oplus \cz$. This naturally split up the reduced configuration space into three
sectors labelled by the kind of vector the $SO(1,2)$ elements stabilize;
time-like,
null or space-like. Only the space-like sector correspond to conventional
geometrodynamics, and if the intention is to compare with {\it e.g} the
ADM-approach,
one chooses this sector. In studying this sector, I choose
to use the choice made by Carlip \cite{carlip1}
for the representation of $\p _1(T^2)$:
\be \L _1:=\left (\begin{array}{ccc} \cosh (\l) & \sinh (\l ) & 0 \\ \sinh (\l
) &
\cosh (\l ) & 0 \\ 0 & 0 & 1 \end{array}\right ),\hspace{10mm}
\L _2:=\left (\begin{array}{ccc} \cosh (\m ) & \sinh (\m ) & 0 \\ \sinh (\m ) &
\cosh (\m ) & 0 \\ 0 & 0 & 1 \end{array}\right ) \ee
with the identification $(\l ,\m )\sim -(\l ,\m )$ that follows from a small
gauge
transformation.
Thus the resulting reduced theory is described by two pair of canonically
conjugate
variables and a vanishing Hamiltonian, {\it i.e} the theory is completely
trivial.
In the notation used by Carlip \cite{carlip1}, we have
$\{\m , a\}=\frac{1}{2}$ and $\{\l , b\}=-\frac{1}{2}$ and all other Poisson
brackets
vanish. Here, $a$ and $b$ are the phase space momenta.
The configuration space $\cc $ is topologically the punctured plane
$\car ^2 - (0,0)$ where points reflected
through
the origin are identified. Now, in the transition to the reduced phase space,
we have
taken care of all the small $SO(1,2)$ transformations as well as all the small
diffeomorphisms. However, we still have the large $O(1,2)$ transformations as
well as
the large diffeomorphisms to deal with. These large transformations are
symmetries of
the classical theory, and should therefore presumably also be symmetries of the
quantum theory. In this case, however, it is enough to have the wavefunctions
transforming under a unitary representation of this remaining symmetry group.
The large or global diffeomorphisms on the torus are normally called modular
transformations. Their action on the configuration space is given
by
\bea S: \l \rightarrow \m ,&&\m \rightarrow -\l  \label{S2transf} \\
T:   \l \rightarrow \l ,&& \m \rightarrow \m + \l  \label{T2transf}
\eea
What is left of the large $O(1,2)$ transformations are the internal
time-reversal and
parity transformations. However, as is easily seen, they have trivial action on
the
configuration space, meaning that we do not have to bother about them. Thus, we
only
have the modular transformations left. By representing a point in the
configuration
space as $v:=\left (\stackrel{\m }{\l }\right )$, we may represent the $S$ and
$T$
transformations as \\
\bea S&=&\left ( \begin{array}{cc} 0 & -1 \\ 1 & 0 \end{array} \right ) \\
T&=&\left ( \begin{array}{cc} 1 & 1 \\ 0 & 1 \end{array} \right )
\label{SoT}\eea
which are easily recognized as the generators of the group $SL(2,\cz )$.
However,
since we should identify two points that are related via a reflection through
the
origin $v\sim -v$, it follows that we have to identify two $SL(2,\cz)$ elements
that
differ only by the sign.
Thus, effectively we are dealing with the group
\footnote{Since $PSL(2,\cz )=SL(2,\cz )/\{{\bf
1},{\bf -1}\}$, an element in $PSL(2,\cz)$ is really the equivalence class
$\{G,-G\}$,
$G\in SL(2,\cz )$. However, since our configuration space is such that $v=-v$,
we may
just choose one representative from the class to represent the $PSL(2,\cz )$
element.} $PSL(2,\cz )\simeq
SL(2,\cz )/\{-{\bf 1},{\bf 1}\}$. Using this representation of the generators
of the group, two
important relations follow immediately:
 \be S^2=(S\cdot T)^3=-{\bf 1}\sim {\bf 1}. \label{rels} \ee \\ \\
Now it is time to study some finite dimensional irreducible representations
of our symmetry group $PSL(2,\cz )$ in a Hilbert space. I will work with
two different Hilbert spaces:\\ \\

Case I: $\ch =L^2(\cc ,d\m d\l )$ and every vector in $\ch $ is required
to transform under a $N$-dimensional unitary representation of $PSL(2,\cz
)$ according to
\bea \J (v)&=&\sum _{l=1}^N\sum _{m=1}^{Dim \ch /N} C^{lm}B_{lm}(v) \\
\J(G\cdot v)&=&\sum _{l=1}^N\sum _{n=1}^N\sum _{m=1}^{Dim \ch
/N}C^{lm}D^N(G)_l{}^nB_{nm}(v) \eea
where $B_{lm}(v)$ is an orthogonal basis in $\ch $, $G\in PSL(2,\cz )$, and
$D^N(G)_l{}^n$ is an element of an $N$-dimensional unitary irreducible
matrix-representation of $PSL(2,\cz )$. Note that if $\ch $ is infinite
dimensional we may choose a continuous label $m$ for the "basis" and have to
replace the summation over $m$ by an integral. The inner-product in $\ch$ is:
$<\J _1|\J _2>:=\int _{\cc}d\m d\l \bar{\J }_1\J _2$\\ \\

Case II: $\tilde{\ch}_N=\sum _{\oplus}\ch _s = \ch _1 \oplus \ch _2 \oplus
\cdots \oplus \ch _N$, where each $\ch _s=L^2(\cc ,d\m d\l)$, and I require
the vectors in $\tilde{\ch }_N$ to transform under an $N$-dimensional unitary
representation of $PSL(2,\cz )$ according to
\bea {\bf \J}(v)&:=&\left (\begin{array}{c} \J _1(v) \\ \J _2(v) \\ \cdot \\
\cdot \\ \J _N(v) \end{array}\right ),\hspace{10mm} \J _s(v)\in \ch _s \\
{\bf \J}(G\cdot v)&=&{\bf D}^N(G)\cdot {\bf \J}(v) \eea
where $G\in PSL(2,\cz )$ and ${\bf D}^N(G)$ is an element of an
$N$-dimensional unitary irreducible matrix-representation of $PSL(2,\cz )$.
The inner-product is here given by $<{\bf \J}_1|{\bf \J}_2>:=\int _{\cc}d\m
d\l {\bf \J}_1^\dagger \cdot {\bf \J}_2 $ \\ \\
Now, if we manage to show that $\tilde{\ch }_N$ is empty, we have
automatically also shown that $\ch $ is empty, and it is thus sufficient to
study only case II to prove the non-existence of the finite dimensional
representations. To see that this is indeed the case, suppose that
$\tilde{\ch }_N$ is empty while $\ch $ is not. This means that there exist a
set of orthonormal basis vectors in $\ch $ such that $B_{lm}(G\cdot
v)=\sum _{n=1}^ND^N(G)_l{}^nB_{nm}(v)$. Now, construct the following vector
in $\tilde{\ch }_N$:
\be {\bf \J}_B(v):=\sum_{m=1}^{Dim \ch /N}\left (\begin{array}{c}
f_1^mB_{1m}(v) \\ f_2^mB_{2m}(v) \\ \cdot \\ \cdot \\ f_N^mB_{Nm}(v)
\end{array} \right ) \ee
where the $f_s^m$ are arbitrary Fourier-coefficients.
As is easily checked, this vector transforms correctly to belong to
$\tilde{\ch}_N$ and its norm is
\be <{\bf \J}_B|{\bf \J}_B>=\sum_{s=1}^N\sum_{m=1}^{Dim \ch /N}
\bar{f}_s^mf_s^m \ee
and thus, by properly choosing the $f_s^m$'s we have an element in
$\tilde{\ch }_N$, which is a contradiction. Therefore we can conclude that if
$\tilde{\ch }_N$ is empty, it follows that no basis $B_{lm}(v)$ in $\ch $
with the correct transformation property exists and hence, $\ch $ is empty as
well.\footnote{I gratefully thank Don Marolf for suggesting this proof.}
\\ \\

To see what kinds of restrictions we get on the finite dimensional irreducible
representations, we will
need some results concerning the orbits of the modular transformation on our
configuration space. Basically, I intend to show that all the rational lines --
$\frac{\m }{\l}$ is rational -- are equivalent under the modular
transformations, and
that every point on a rational line is left invariant by an abelian invariant
subgroup of $PSL(2,\cz)$
-- different subgroups for different rational lines.
Moreover, I will show that it is possible to transform two different points on
different rational lines arbitrarily close to different points on the same
rational
line.

These three results are enough to put extremely hard restrictions on the
admissible
representations as well as wavefunctions. Since every point on a rational line
is left
invariant by the action of an element -- the generator of the abelian
subgroup that stabilizes the line -- of $PSL(2,\cz)$, we have to require the
wavefunctions to be eigenvectors with eigenvalue one to the representation of
the
element under which the line is left invariant. That is, we have to require our
representation of $PSL(2,\cz)$ to be such that the representation of the group
elements
that have fixpoints all have the eigenvalue one in its spectrum. And since
there are
an infinite number of such elements in $PSL(2,\cz)$ this seems to be a severe
restriction on what representations that are allowed. Furthermore, even if such
non-trivial
representations may be found we still have to
require the wavefunctions to be continuously extendable outside the rational
lines.
Actually, in the  finite dimensional case it is enough to require the
wavefunctions to be piecewise continuous somewhere to rule out all
representations besides the trivial
one, and for that representation we can only allow the constant wavefunction.

To simplify the transformations, I choose my configuration
space to be defined as $\l \geq 0$, $\m \in \car$ with the identification $(\l
,\m
)=(0, \m )\sim (0, -\m )$ and define the new
coordinates
\be y:=\frac{\m}{\l},\hspace{10mm}x:=\l. \label{nykoord} \ee
In these coordinates, the modular transformations become
\bea S: y\rightarrow -\frac{1}{y},&& x\rightarrow x\; y \label{S5transf} \\
T: y\rightarrow y+1,&& x\rightarrow x \label{T5transf} \eea
and $v=\left (\begin{array}{c} x\;y\\x\end{array}\right )$. Throughout the rest
of the
paper, I will use three different notations to denote a point in $\cc$: $v$,
$(\cdot
,\cdot )_{xy}$ or $(\cdot ,\cdot )_{\l \m}$.
Consider now the transformation $V:=T^2\cdot S$ where $T^2:=T\cdot T$.
Under multiple applications of $V$
on a generic starting point $v=v_0$, we get $v_n=V^n\cdot v_0$:
\bea \left (\begin{array}{c}x_n\; y_n \\x_n \end{array} \right )=
\left (\begin{array}{cc} 2
& -1 \\ 1 & 0 \end{array}\right )^n\cdot \left (\begin{array}{c}x_0\; y_0 \\x_0
\end{array} \right )= \nn \\ \left (\begin{array}{cc} (n+1)
& -n \\ n & -(n-1) \end{array}\right )\cdot \left (\begin{array}{c}x_0\; y_0
\\x_0
\end{array} \right )=\left (\begin{array}{c} x_0(y_0(1+n)-n)\\ x_0(n\; y_0
+(1-n))
\end{array}\right ). \label{xnyn} \eea
We can directly read off $x_n=x_0(n\; y_0 +(1-n))$ and $y_n=\frac{y_0(1+n)-n}
{n\; y_0 +(1-n)}$. Thus for $n\gg 1$ and $|n(y_0-1)| \gg 1$ we get $y_n\sim 1+
\frac{1}{n} +
\co (\frac{1}{n^2(y_0-1)})$, and we see that we can map a generic starting
point
$x_0,y_0$ arbitrarily close to the line $y=1$.
 For rational
$y$ it is actually possible to prove an even stronger statement:\\ \\
Theorem: All rational lines $y=\frac{p}{q}$, $p,q\in \cz$ are equivalent under
$PSL(2,\cz )$ transformations. \\ \\
Proof: Consider a point on the line $y=\frac{m}{n}$, $m,n \in \cz $
 and $g.c.d(m,n)=1$. Here, $g.c.d(\cdot, \cdot )$ denotes the greatest common
divisor of the entries. First of all, we may note that all points on rational
lines are of the above form.

To see that the line $y=\frac{m}{n}$ always can be mapped into the line $y=0$,
consider
\be \left (\begin{array}{c} 0 \\ x' \end{array} \right ) = \left
(\begin{array}{cc}
N_1 & N_2 \\ N_3 & N_4 \end{array} \right )\cdot \left (\begin{array}{c} x\;
\frac{m}{n} \\ x \end{array}\right ) = \frac{x}{n}\left (\begin{array}{c}
mN_1+nN_2 \\
m N_3+n N_4 \end{array} \right ) \ee
where $N_1,N_2,N_3,N_4\in \cz $ and $N_1N_4-N_2N_3=1$. Thus we see that
$N_1=p\;n$ and
$N_2=-p\; m$, $p\in \cz$. Then, we just need to make sure that there exist
integers
$N_3$ and $N_4$ such that $p\; n\; N_4+p\; m\; N_3=1$. However, as is well
known
\cite{number}, this equation has a solution iff $g.c.d(p\; n,p\; m)=1$
$\Leftrightarrow
p=\pm 1$, and hence $x'=\pm \frac{x}{n}$. Thus, we get the result
\be (x,\frac{m}{n})_{xy}\stackrel{SL(2,\cz
)}{\leftrightarrow}(\frac{x}{n},0)_{xy}
\label{conj} \ee
for $x\in \car $ and $m,n\in \cz $ and $g.c.d(m,n)=1$. This proves the
theorem. \\ \\
Thus, we now know that all points on rational lines are equivalent
under the modular transformations. Furthermore, for irrational lines, we
have seen that it is possible to transform us arbitrarily close to the line
$y=1$ by
repeatedly applying modular transformations. With this result at hand it is
rather
obvious that there does not exist a fundamental region for the $PSL(2,\cz )$
transformations in the configuration space. Remember that the conventional
definition
of a fundamental region \cite{riemm} says that the interior of the
fundamental region should contain
one and only one representative of points equivalent under the symmetry
transformation. (If one includes the total boundary of the fundamental region,
one has
to allow points on the boundary to be equivalent.) In our case, it seems
rather natural to require the fundamental region to contain one and
only one rational line. However, there are no such connected open regions. As
soon as we try to go outside the rational line by appending a tiny area to
the rational line, we are automatically including segments from an infinite
number of rational lines and these segments can all be mapped to the original
rational line, meaning that there does not exist any fundamental region in
the conventional sense.

A direct consequence of the absence of a fundamental region is that
no wavefunctions that transform under a finite dimensional unitary
representation of
$PSL(2,\cz )$ will be normalizable w.r.t the inner-product that is defined over
the
entire configuration space. That is, $\tilde{\ch }_N$ is empty and
consequently so is $\ch $!
To see this, consider the inner-product in $\tilde{\ch}_N$: $<{\bf
\J}_1|{\bf \J}_2>=\int _\cc d\m d\l {\bf
\J}_1^\dagger \cdot {\bf \J}_2 $, and
 split up the configuration space into an infinite number of regions
equivalent under $PSL(2,\cz )$ transformations: $\cc =\cup _{k\in \cz} W_k$
where
$W_k$ is defined as: $x>0$ and $k\leq y < k+1$. Since $W_k$ is mapped into
$W_{k+1}$
by a $T$ transformation, we see that all $W_k$ regions are equivalent under $T$
transformations. But, since our wavefunction just transform under a finite
dimensional unitary
matrix-representation of $PSL(2,\cz)$, we directly see that the inner-product
becomes:
\be <{\bf
\J}_1|{\bf \J}_2>=\int _\cc d\m d\l {\bf
\J}_1^\dagger \cdot {\bf \J}_2=\infty \times \int _{W_0}d\m d\l {\bf
\J}_1^\dagger \cdot {\bf \J}_2 \ee
Consequently, the inner-product diverges for all ${\bf \J}$ and $\tilde{\ch
}_N$ is thus empty. Note also that the
result would have been the same for any finite dimensional representation,
reducible
or irreducible. If the
representation is infinite dimensional, however, one needs to make sure that
the
infinite scalar product ${\bf
\J}_1^\dagger \cdot {\bf \J}_2$ in the redefinition of the inner-product always
converges. This result -- the non-existence of finite dimensional
$SL(2,\cz )$-invariant
closed subspaces of $L^2(\car ^2,dx dy)$ --
 has recently been proven for $\ch $ with the use of
representation theory for $SL(2,\car )$ in $L^2(\car ^2,dx dy)$ \cite{Jorma}.

This
result is of course nothing new and especially strange; we have the same
situation for
all theories having symmetries that cut up the configuration space into an
infinite number of non-zero area pieces. For example, one may think of
wavefunctions
on $\car$ that transform under a finite dimensional representation of the
$a$-translation group; translations with a fix $a$. However, in this case we
have the
natural option of redefining our inner-product to be integrated only over the
fundamental region : $0\leq x <a$.

In our case, since there is no fundamental region, there is no natural
redefinition of
the inner-product. One could take the drastic step to define the
inner-product to be integrated only over a rational line. However, by doing so,
one is
also drastically changing the Hilbert space; a rational line is a region of
measure zero in the full configuration space. Perhaps a better choice is to
define the
inner-product over a finite area
-- w.r.t the Euclidean metric on $\cc$ -- region in $\cc$. I will, however, not
make
such a choice here. I just assume that one can find a nice, well defined
inner-product
for wavefunctions transforming under a finite dimensional unitary
representation, and
make appropriate changes in the previous definition of the inner-product. \\ \\

Now, returning to the study of the orbits of the modular transformations on
$\cc$, I
intend to show that every point on a rational line is a fixpoint under the
action of
one element of $PSL(2,\cz )$. A generic point on a rational line is given by
\be v_{p,q}:=\left (\begin{array}{c} x \; \frac{p}{q} \\ x \end{array} \right )
\ee
where $p,q\in \cz $, $x\in \car $. This point is left invariant by the
$SL(2,\cz )$ element
\be G_{p,q}:=\left (\begin{array}{cc} 1-p\; q & p^2 \\ -q^2 & 1+p\; q
\end{array}\right
). \label{Gpq} \ee
Actually, it is left invariant by all elements of the form $G_{\a p,\a q}$ for
$\a ^2
\in \cz$. However, these elements are generated by $G_{p,q}$ as $G_{\a p,
\a q}=G_{p,q}^{\a ^2}$ and are thus not important here.

The next result I will need is the fact that two points on different rational
lines
may be mapped arbitrarily close to {\it different} points on the same rational
line.
Consider points of the form
\be v_{(\l ,\a ,p,q,k)}:=\l\; \a\left (\begin{array}{c}
\frac{p}{q}(1-\frac{1}{pqk}(\frac{1}{\a}-1)) \\ 1
\end{array}\right ) \ee
where $p,q,k\in \cz $ and $\l ,\a \in \car $. This line $0<\l \; \a <\infty
$ has the property that it approaches the line $y=\frac{p}{q}$ as
$k\rightarrow \infty$. By using a transformation of the form
$G_{p,q}^k$ -- $G_{p,q}$ raised to the power $k$ --
 this point is mapped into $v_{(\frac{\l}{\a} ,\a,
p,q,\frac{k}{\a})}=G_{p,q}^k\cdot v_{(\l ,\a ,p,q,k)}$:
\be
\l \left (\begin{array}{c} \frac{p}{q}(1-\frac{\a}{pqk}(\frac{1}{\a}-1))
 \\ 1 \end{array}\right )=
\left (\begin{array}{cc} 1-k\;p\;q
& k\;p^2 \\ -k\; q^2 & 1+k\; p\; q \end{array}\right )\cdot
 \l\; \a\left (\begin{array}{c}
\frac{p}{q}(1-\frac{1}{pqk}(\frac{1}{\a}-1)) \\ 1
\end{array}\right )\label{rel8} \ee
which in the limit $k\rightarrow \infty$ says $(\l ,\frac{p}{q})_{xy}
\sim (\l \; \a ,\frac{p}{q})_{xy}$. Or in
words; the points $\l $ and $\l \; \a $ on the line $y=\frac{p}{q}$
is approximately related by an
$PSL(2,\cz)$ transformation for all $\l , \a \in \car $. Note, however,
that this relation never is exact. One can actually easily prove that no two
points on
the same rational line are equivalent under a $PSL(2,\cz )$
transformation\footnote{To see this, try to map a point on the line $y=0$ to a
different point on the same line: $\left (\begin{array}{c} 0 \\ x' \end{array}
\right
)=\left (\begin{array}{cc} N_1 & N_2 \\ N_3 & N_4 \end{array} \right )\cdot
\left (
\begin{array}{c} 0 \\ x \end{array} \right
) =\left (
\begin{array}{c} N_2x \\ N_4x \end{array} \right
)$, $x'\neq x$. We get $N_2=0$ and $N_4\neq 1$. But the only element of
$SL(2,\cz )$
satisfying this is $G=\left (\begin{array}{cc} -1 & 0 \\ N_3 & -1 \end{array}
\right )$ which maps $x\rightarrow -x$ on the line $y=0$. and since these
points are
identified in our configuration space, we see that no two point on the line
$y=0$ are
related by a $PSL(2,\cz )$ transformation. Furthermore, since every point on a
rational line has a unique image on the line $y=0$, the result applies to all
rational lines.}. For
continuous wavefunctions, this approximate relation is, however, good enough.\\
\\
Now, what does all this mean for a wavefunction transforming under a finite
dimensional representation of $PSL(2,\cz)$? First of all, since every point on
a
rational line is a fixpoint for an element of $PSL(2,\cz)$, we have to require:
\be {\bf \J }(v_{p,q})={\bf \J }(G_{p,q}\cdot v_{p,q})={\bf
D}^N(G_{p,q})\cdot {\bf \J }(v_{p,q}) \ee
Hence, the wavefunction is either zero on the line $y=\frac{p}{q}$ or an
eigenvector
with eigenvalue one to ${\bf D}^N(G_{p,q})$. However, if the wavefunction is
identically zero on one rational
line it follows that it must be identically zero on all rational lines.
(Remember that
all rational lines are equivalent under $PSL(2,\cz)$ transformations.) Thus,
non-trivial piecewise continuous wavefunctions must be eigenvectors with
eigenvalue one to
${\bf D}^N(G_{p,q})$. And this means that we can only allow representations
where
${\bf D}^N(G_{p,q})$ has the eigenvalue one in its spectrum for {\it all} $p$
and $q$!
Such representations exists. One example is the trivial representation; all
elements
in $PSL(2,\cz )$ are represented by the identity matrix.
 It may, however, be impossible to find a faithful
finite dimensional unitary representation that allows this.

Suppose now that we have found such a representation and we want to realize
it in the space of piecewise continuous complex valued $N$-dimensional
vector-functions on $\cc$. Now, since we require the functions to transform as
${\bf \J}(G\cdot v)={\bf D}^N(G)\cdot {\bf \J}(v)$ eq. (\ref{rel8}) says
\be {\bf \J}(G^k_{p,q}\cdot v_{(\l ,\a ,p,q,k)})=\left ({\bf
D}^N(G_{p,q})\right )^k\cdot {\bf \J}(v_{(\frac{\l}{\a} ,\a,
p,q,\frac{k}{\a})} ) \ee
which in the limit $k\rightarrow \infty$ becomes
\be {\bf \J}(\l , \frac{p}{q})_{xy}={\bf \J}(\l \; \a ,\frac{p}{q})_{xy}
\label{rel10b}\ee
where I have used that ${\bf \J}(x,\frac{p}{q})_{xy}$ is stable under ${\bf
D}^N(G_{p,q})$, and that entries in unitary matrices have modulus $\leq$1.
This makes the limit well defined.

Thus, since eq.(\ref{rel10b}) is true for all $\l ,\a \in \car$, we may
conclude
that the wavefunction is constant on every rational line, and hence it is
constant everywhere in $\cc$.

This completes the proof of the nonexistence of nontrivial piecewise continuous
wavefunctions transforming under a unitary finite dimensional representation
of $PSL(2,\cz )$\footnote{In
\cite{carlip1} there is a claim that wavefunctions transforming under a
one-dimensional nontrivial representation of $SL(2,\cz)$ are being constructed.
However, as was already noted in \cite{louko}, there is a subtlety in the
integration
region in eq. (3.3) in ref. \cite{carlip1}.
If the integration region is chosen to be the moduli space, the
wavefunctions fail to transform correctly, and if the integration region is the
entire Teichm\"{u}ller space, the integral is illdefined and diverges.}. \\ \\

Before leaving this representation, I want to comment on possible ways of
fixing the
problem of using finite dimensional representations. If we want to construct
our
Hilbert space from the vector space of piecewise continuous
complex valued functions on $\cc$, there
are no good solutions. That is, since already this underlying vector space does
not
 allow finite dimensional irreducible representations, there is no way to fix
our
problem by making a clever choice of an inner-product. However, if we are
prepared to
relax the smoothness of the functions to only requiring continuity along the
rational
lines, it may be possible to use finite dimensional representations. That
is, we will get wavefunctions that are nowhere continuous in a two-dimensional
neighborhood of a point in $\cc$. Our inner-product could be defined as the
integral
over one rational line. This situation will very much resemble the case for the
loop-representation, discussed below. However, one should be aware of that this
construction is very different from the conventional Hilbert space
constructions in
standard quantum mechanics. Normally, although one sometimes restricts the
integration
region to a fundamental region, there is no problem to extend the wavefunctions
smoothly outside this fundamental region to the entire configuration space.
This will
not be possible here. Also, with such a Hilbert space, it is not clear how to
find a
quantum representation of {\it e.g} the classical Poisson algebra for the phase
space
variables.

Another, perhaps more promising way to fix the problem is to choose a different
polarization of the phase space. Since we know that there is a one-to-one
mapping
between the ADM-phase space and half the connection phase space, and there
exist a
fundamental region in the ADM phase space, there must exist a fundamental
region in
the connection phase space \cite{carlip1}.
I have only shown that this fundamental region "has a
non-trivial projection" down to the configuration space. Thus, by choosing a
different
polarization it may be possible to use a conventional Hilbert space to find a
non-trivial quantum theory even for these sectors\footnote{I thank Abhay
Ashtekar for
suggesting this.}.However, such a quantum representation is strictly speaking
not a
connection representation. Results of such a quantization will be reported
elsewhere.
\section{Loop-representation}
There is a closely related formulation of quantum gravity called the
loop-representation \cite{ashtekar1}, \cite{marolf}.
In that formulation, one starts from the same
Hamiltonian formulation as for the Chern-Simons/connection formulation, but
instead of
quantizing the classical Poisson algebra of the reduced phase space variables,
one
uses a classical T-algebra of holonomy-like variables $T^0$, $T^1$
to coordinatize the phase space.
By using these variables which are manifestly gauge invariant, one has
automatically taken care of the local gauge invariance. The local
diffeomorphism
invariance is then handled by restricting the theory to flat connections.
In the quantization of this algebra, one defines a quantum representation in
the
loop-space of the hypersurface. And, basically due to the fact that one uses
flat
connections, these $T$ operators will only be able to distinguish between
homotopically inequivalent loops. On the torus, every homotopy class of loops
is uniquely labelled by two integers; the winding numbers around the two
noncontractible independent loops on the torus. Thus, all information in a
function on
loop-space as seen by the $T$-operators are given by these two integers.
Effectively,
this means that the Hilbert space will be based on the vector space of complex
valued
functions on the two-dimensional lattice $\cz ^2$. Actually, there is one
additional
degeneracy; due to classical relations among the T-variables, the quantum
 representation is only well defined on wavefunctions satisfying
$\J(n_1,n_2)=\J
(-n_1,-n_2)$. Effectively this
means that we should identify two points on the lattice that are related via a
reflection in the origin: $(n_1,n_2)\sim-(n_1,n_2)$. What is left of the global
transformations is again the modular transformations of the torus; they are
given by
\cite{carlip2}
\bea S: &&(n_1,n_2)\rightarrow (n_2,-n_1) \\
T: &&(n_1,n_2)\rightarrow (n_1,n_1+n_2) \eea
Thus, basically we are now dealing with the lattice formulation of the
configuration
space for the Chern-Simons/connection representation.\\ \\
One could now try to go through the same analysis of the irreducible sectors of
this
quantum theory as was done for the connection representation. In this case,
however,
we do not find such strong statements about the finite dimensional irreducible
representations. Basically, this comes about because the underlying space is
discrete;
we cannot use continuity argument anymore. We still have the problem of finding
representations where all ${\bf D}^N(G_{p,q})$ has the eigenvalue one in
its spectrum, but once this is done, there should not be any problem of finding
non-trivial wavefunctions transforming under this representation. As
an example of
allowed representations we have the trivial one-dimensional representation.

Note also that here as well we get the problem that a
wavefunction transforming under a finite dimensional representation will not be
normalizable over the entire lattice. That is, if we define the inner-product
to be
\be <\J _1|\J _2>:=\sum _{n_1,n_2}\bar{\J}_1(n_1,n_2)\J _2(n_1,n_2) \ee
we automatically exclude wavefunctions transforming under a finite dimensional
representation of $PSL(2,\cz)$. This can be understood from the fact that the
two-dimensional lattice can be written as the infinite sum of all rational
lines, and
that all rational lines are equivalent under $PSL(2,\cz)$ transformations. And
if one
tries to modify the inner-product by restricting the summation to only one
rational
line, the $T$-operators as defined in \cite{ashtekar1} will not be self
adjoint. Thus,
it is straightforward to find a non-trivial vector space of {\it e.g} modular
invariant wavefunctions here, but we still do not have a good inner-product
that makes
these wavefunctions normalizable and makes the $T$-operators self adjoint.

\section{The reduced ADM-phase space approach}
In the ADM-approach to quantum gravity, one starts from the ADM-Hamiltonian
formulation of gravity and reduces the phase space classically by solving all
first
class constraints and gauge fixes the local symmetries they generate
\cite{ADM},
\cite{carlip1}, \cite{carlip2}, \cite{hosoya}. In
the torus case, this can be done completely explicetely; the reduced
configuration
space is parametrized by the Teichm\"{u}ller coordinates for the torus, and we
get the
Hamiltonian
\be \hat{H}=\sqrt{\t _2^2(\frac{\partial ^2}{\partial \t _1^2}+
\frac{\partial ^2}{\partial \t _2^2})} \ee
where $\t _1$ and $\t _2$ are the coordinates on the Teichm\"{u}ller space.
What is
left of the global symmetries are again the modular transformations, whose
action on
the configuration space is
\bea S:&&\t\rightarrow -\frac{1}{\t} \\
T:&&\t \rightarrow \t +1 \eea
where $\t:=\t _1 +i \t _2$. These transformations on this space are well
studied in
the mathematics
literature \cite{riemm}, \cite{maass} and it is known that we here get a
fundamental region:
$\cm$:
$|\t |\geq 1$ and $|\t _1|\leq \frac{1}{2}$. The Teichm\"{u}ller space is
naturally
cut up into an infinite number of copies of this fundamental region. This means
that we again has to choose integration region in our inner-product with care.
Choosing an integration over the entire Teichm\"{u}ller space automatically
excludes
all wavefunctions transforming under a finite dimensional representation of
$PSL(2,\cz
)$. Therefore, since we here have a fundamental region, it seems more natural
to
define the inner-product over the moduli space $\cm$. There also exist a
natural
metric, and hence a volume element, on the Teichm\"{u}ller space; the
Poincar\'{e}
metric $ds^2=\t _2^{-2}(d\t _1^2 + d\t _2^2)$ with volume element
$\sqrt{g}d\t _1d\t _2=\t _2^{-2}d\t _1d\t _2$.
The Hamiltonian given above is the square
root of the Laplacian w.r.t this metric, and it may be shown that the classical
solutions to this theory are given by geodesics w.r.t this metric \cite{ADM}.
Thus,
it seems natural to choose the inner-product
\be <\J _1|\J _2>:=\int _\cm \frac{d\t _1d\t _2}{\t_2^2}\bar{\J }_1\J _2 \ee
I do not know what can be said about the generic irreducible sector in this
Hilbert
space, it is, however, known that the trivial one-dimensional unitary
representation
-- the space of moduli invariant wavefunctions -- here is non-trivial. That is,
this
sector is known to infinite dimensional and mathematicians have numerically
studied a
basis of eigenfunctions to the Laplacian; the so called zero
weight Maass-forms \cite{maass},
\cite{modforms}.
\section{Conclusions}
In this paper, I have shown that the connection representation
for (2+1)-quantum gravity on the torus is completely trivial if
we require the wavefunctions to transform under a
finite dimensional representation of $PSL(2,\cz)$, and also require
the wavefunctions to be piecewise continuous somewhere on the
reduced configuration space. Thus if we strongly believe that
the wavefunctions must transform according to a one-dimensional
representation of $Diff(T^2)/Diff(T^2)_0$ and want a nontrivial
quantum theory, we are forced to accept a strangely looking
Hilbert space; it will consist of wavefunctions that are nowhere
continuous. And thus, since no one yet has shown how to find an commutator
representation
of some classical poisson-algebra in such a Hilbert space,  we are yet far from
having
constructed a sensible nontrivial quantum theory in this sector. This
immediately also implies problems for the loop-representation.
According to \cite{ashtekar1}, we probably need to relate the
loop-representation to the connection representation via the
loop-transform in order to be sure that our Hilbert space will
be properly constructed as to contain the conventional
geometrodynamical sector.

What is a bit puzzling here, is that we already have a
nontrivial quantum theory in the trivial representation,
for the torus case; the ADM-representation \cite{hosoya}. This
seems to indicate that either the connection- and
ADM-representation are not isomorphic/unitarily equivalent
\underline{or} that the ADM-representation is unitarily
equivalent to a connection representation where our
wavefunctions has to be strangely looking objects (nowhere
continuous).

We should also remember that it may still be possible to find a nontrivial
quantum
theory with nice wavefunctions coming from the connection formulation if we use
a
different polarization of the phase space.\\ \\
At last, note that there already exist a nontrivial quantum theory
for (2+1)-quantum gravity on the torus in the connection representation if we
are
prepared to accept that the wavefunctions transform under an infinite
dimensional,
reducible
representation of $PSL(2,\cz)$ \cite{ashtekar1}, \cite{marolf}, \cite{Jorma}.
\\ \\
\underline{Acknowledgements}\\
I thank Arlen Anderson, Abhay Ashtekar, Fernando Barbero, Ingemar Bengtsson,
Domenico Giulini, Jorma Louko and Don Marolf for
interesting discussions. I also thank Nigel Higson and Jean-Luc Brylinski for
some
information regarding representation theory for $SL(2,\cz )$.
This work was supported by the NFR contract F-PD 10070-300.

\newpage \pagestyle{plain}

\end{document}